\documentstyle[pre,aps,multicol,epsfig,amstex]{revtex}

\begin{document}
\draft
\preprint{MA/UC3M/05/1998}
\title{Phase behaviour of additive binary mixtures in the limit of 
infinite asymmetry}
\author{Yuri Mart\'\i nez-Rat\'on and Jos\'e A.\ Cuesta}
\address{Grupo Interdisciplinar de Sistemas Complicados
(GISC), Departamento de Matem\'aticas, Escuela Polit\'ecnica
Superior, Universidad Carlos III de Madrid, c/ Butarque, 15, 28911 --
Legan\'es, Madrid, Spain}

\maketitle

\begin{abstract}
We provide an exact mapping between the density functional of a
binary mixture and that of the effective one-component fluid in the
limit of infinite asymmetry. The fluid of parallel hard cubes is thus
mapped onto that of parallel adhesive hard cubes. Its phase
behaviour reveals that demixing of a very asymmetric mixture can
only occur between a solvent-rich fluid and a permeated large particle
solid or between two large particle solids with different packing
fractions. Comparing with hard spheres mixtures we conclude
that the phase behaviour of very asymmetric hard-particle mixtures
can be determined from that of the large component interacting via
an adhesive-like potential.
\end{abstract}

\begin{multicols}{2}
\narrowtext
\tighten

Since Biben and Hansen showed \cite{biben0} that a binary mixture of
hard spheres (HS) in Rogers-Young approximation has a spinodal
instability for diameter ratios smaller than 0.2, it is generally
accepted that a demixing transition occurs in this fluid if the 
asymmetry between the two component sizes is large enough. This
fact has been confirmed with other theories
\cite{lekkerkerker}, simulations \cite{buhot}, 
and experiments in colloidal suspensions 
\cite{dinsmore}. The question is no more
whether such a mixture demixes but which is its phase behaviour.
Some approximate theories \cite{poon} suggest that the
phase rich in large spheres is a crystal instead of a fluid, in
agreement with what is observed in the experiments
\cite{dinsmore}. The depletion potential between the
large spheres due to the small ones has been determined both
perturbatively \cite{mao} and in simulations \cite{biben3}. If the
interaction of the small spheres is
replaced by this potential the resulting effective
fluid can be simulated \cite{biben3,noe,dijkstra}. In the unstable
region of the binary mixture large clusters of depleted spheres are formed,
which evolve very slowly \cite{biben3}. The phase behavior of this
effective fluid for diameter ratios smaller than 0.1 \cite{noe,dijkstra}
shows a large coexistence region between a diluted fluid and a highly
packed solid, and even an isostructural solid-solid transition, if
the small spheres packing fraction and the diameter ratio are low enough.

The picture emerging from these studies is that the phase behaviour
of a very asymmetric binary mixture can be understood in terms of the
phase behaviour of a fluid of HS interacting via a depletion potential.
This potential has a very narrow (essentially the diameter of the small
spheres) and deep well \cite{biben3,mao}, and this type of potential
have been proven to have no liquid phase \cite{bolhuis,tejeroetal}:
there is only fluid-solid coexistence, or---if it is narrower and
deeper---in addition there may appear an isostructural solid-solid
transition.

In this letter we will explore this connection between the phase
behaviour of an additive binary mixture and that of the effective
depleted one-component fluid by explicitly showing that the Helmholtz
free-energy functional of the mixture reduces to that of the 
corresponding one-component adhesive fluid in the limit of infinite
asymmetry. That in this limit the solvent induces an adhesive
interaction between the large particles is already a known result
\cite{biben,heno}; but we are going a step further by mapping the
{\em functional} of the mixture onto that of the corresponding
adhesive fluid, and hence their respective phase behaviours.

The procedure is as follows. As usual in these systems we fix the
chemical potential of the small particles, $\mu_S$ (semigrand ensemble),
and therefore the system is described by the functional
\begin{equation}
\Upsilon(\mu_S,[\rho_L])=F[\rho_L,\rho_S]-\mu_S\int\rho_S, \quad
\mu_S=\frac{\delta F}{\delta\rho_S({\bf r})},
\label{upsilon}
\end{equation}
$\rho_{L(S)}$ being the density profile of the large (small) particles,
and $F$ the Helmholtz free-energy functional. The equilibrium equation
for $\rho_S$ in (\ref{upsilon}) is
usually a nonlinear equation involving both densities; the infinite
asymmetry limit then allows to explicitly obtain $\rho_S$ as an
expansion with terms depending on $\rho_L$. This operation transforms
(\ref{upsilon}) into an effective Helmholtz free-energy functional for
the depleted large-component fluid (up to some divergent terms which
do not affect the phase behaviour), with the fugacity of the solvent
playing the role of an inverse temperature.

The direct correlation function (DCF) of the effective fluid is defined
as minus the second functional derivative of the excess part of
(\ref{upsilon}) with respect to $\rho_L$. It is straightforward to see
that this leads to the following relationship with the components of
the DCF of the mixture, written in Fourier space:
\begin{equation}
\widehat{C}_{\rm eff}(k)=\widehat{C}_{LL}(k)+\frac{\rho_S
\widehat{C}_{LS}(k)^2}{1-\rho_S\widehat{C}_{SS}(k)}.
\label{effDCF}
\end{equation}
In terms of the structure factor this expression is equivalent to
assuming $S_{\rm eff}(k)=S_{LL}(k)$. This is precisely the procedure
used in \onlinecite{heno} to map a binary HS mixture onto an adhesive
HS fluid in the limit $\epsilon\equiv\sigma_S/\sigma_L\to 0$,
$\sigma_{L(S)}$ denoting the diameter of the large (small) spheres.

To proceed with the functional mapping we have described above we need
a free-energy functional able to deal with strong inhomogeneities (the
solvent can be regarded as a fluid confined in between the large 
particles). Fundamental measure functionals are perfect candidates
\cite{rosenfeld,rosenfeldetal,cuesta,cuesta2} both because
of their ability to reduce dimensionality
\cite{rosenfeldetal,cuesta2} and because they are formulated
directly for mixtures. On the other hand, for the sake of simplicity
we have chosen to apply this procedure to a binary mixture of parallel
hard cubes (PHC).
This fluid has also been shown to have a spinodal
instability both in simulations on a lattice \cite{dijkstra2} and in
continuum space \cite{cuesta}, thus its phase behaviour can be expected
to be qualitatively similar to that of HS. The procedure applied to
the latter, however, has a higher level of complexity both because
of the more complicated shape of the functionals
\cite{rosenfeldetal} and because the weaker depletion in this
fluid requires to expand to a higher order in $\epsilon$. There is 
nothing fundamental, though, which prevents its application to HS, and
we are currently working along this line.

In order to determine the appropriate scaling of the solvent packing
fraction $\eta_S$ we first take the $\epsilon\to 0$ limit in
(\ref{effDCF}) forcing it to produce a finite nontrivial result. The
right choice is $\eta_S=O(\epsilon)$; thus, if we define 
$\eta_S=\epsilon\xi$, 
the resulting DCF will be ($\sigma_L=1$ in what follows)
$C_{\rm eff}=C_{\rm PHC}+C_{\rm ad}$, where
\begin{equation}
C_{\rm ad}({\bf r})=\frac{z}{2(1-\eta)}
\{\delta_{\rm cont}({\bf r})+yS({\bf r})+6y^2V({\bf r})\};
\label{PAHC-DCF}
\end{equation}
$C_{\rm PHC}({\bf r})$ is the DCF of a one-component PHC fluid
\cite{cuesta,cuesta2} of packing fraction
$\eta=\eta_L$ (actually the total packing fraction in the
limiting fluid); $z\equiv\epsilon^2\exp(\beta\mu^{\rm ex})=\xi/(1-\eta)$
is the ``renormalized'' excess (over ideal) fugacity of the solvent;
$y\equiv\eta/(1-\eta)$; and
\begin{eqnarray}
\delta_{\rm cont}({\bf r}) &=& \delta(1-|x|)L(y)L(z)+\delta(1-|y|)L(x)L(z)
\nonumber \\
 &&+\delta(1-|z|)L(x)L(y)  \, ,
\label{deltaS}  \\
S({\bf r}) &=& \Theta(1-|x|)L(y)L(z)+\Theta(1-|y|)L(x)L(z)
\nonumber \\
 &&+\Theta(1-|z|)L(x)L(y)  \, ,
\label{surface}  \\
V({\bf r}) &=& L(x)L(y)L(z)  \, ,
\label{volume} 
\end{eqnarray}
with $L(u)\equiv(1-|u|)\Theta(1-|u|)$ and $\Theta(v)=1$ if $v\geq 0$
and $=0$ otherwise.

Apart from corrections proportional to the overlap surface
(\ref{surface}) and volume (\ref{volume}), at larger densities, the
lowest density correction
(\ref{deltaS}) is a delta on the contact surface---also
proportional to the actual contact surface. This has its origin in
the effective {\em adhesive} potential on the surfaces of the cubes
induced by the solvent, and it turns out to be a natural extension
of that for HS \cite{heno,baxter}. We will thus refer to this fluid 
as the parallel adhesive hard cube (PAHC) fluid.

Once we know the scaling of the density we can determine the free
energy with the procedure described above. The fundamental measure
functional of a mixture of PHC is given by \cite{cuesta2}
$\beta F=\int d{\bf r}\,\{\Phi_{\rm id}({\bf r})
+\Phi_{\rm ex}({\bf r})\}$, where
\begin{eqnarray}
\Phi_{\rm id}({\bf r}) &=&
\sum_i\rho_i({\bf r})\Bigl(\log[{\cal V}_i\rho_i({\bf r})]-1\Bigr) \, ,
\label{ideal}  \\
\Phi_{\rm ex}({\bf r}) &=& -n_0\ln(1-n_3)+
   \frac{{\bf n}_1\cdot{\bf n}_2}{1-n_3}+\frac{n_{2,x}n_{2,y}n_{2,z}}
   {(1-n_3)^2},
\label{excess}
\end{eqnarray}
${\cal V}_i$ being the thermal volume of species $i$, $\rho_i({\bf r})$
its local number density, and $n_0$, ${\bf n}_1$, ${\bf n}_2$, $n_3$ are
a set of ``fundamental'' weighted densities generically defined as
the convolution $n_{\alpha}=\sum_i\rho_i*\omega^{(\alpha)}_i$,
where
$\omega^{(0)}_i\equiv\delta_i^x\delta_i^y\delta_i^z$,
${\bf w}^{(1)}_i\equiv
     (\theta_i^x\delta_i^y\delta_i^z,\delta_i^x\theta_i^y\delta_i^z,
     \delta_i^x\delta_i^y\theta_i^z)$,
${\bf w}^{(2)}_i\equiv
     (\delta_i^x\theta_i^y\theta_i^z,\theta_i^x\delta_i^y\theta_i^z,
     \theta_i^x\theta_i^y\delta_i^z)$, 
$\omega^{(3)}_i\equiv\theta_i^x\theta_i^y\theta_i^z$,
with $\theta_i^u=\Theta(\sigma_i/2-|u|)$, and
$\delta_i^u=(1/2)\delta(\sigma_i/2-|u|)$.

According to (\ref{upsilon}) the equilibrium scaled density of the
solvent $\xi({\bf r})\equiv\epsilon^2\rho_S({\bf r})$ is given by
\begin{equation}
\log\xi({\bf r})=\beta\mu^{\rm ex}_S+\ln\epsilon^2-\sum_{\alpha}
\frac{\partial\Phi_{\rm ex}}{\partial n_{\alpha}}*
\omega_S^{(\alpha)}({\bf r}).
\label{equil}
\end{equation}
If we now take into account that, for any function $f({\bf r})$,
$f*\omega_S^{(\alpha)}=\epsilon^{\alpha}f+
O(\epsilon^{\alpha+2})$, for $\alpha=3$ or any vector component of
$\alpha=2,1$, and $f*\omega_S^{(0)}=f+(\epsilon^2/8)\nabla^2 f+
O(\epsilon^4)$, then
$\xi({\bf r})=z[1-\overline{n}_3({\bf r})]+\epsilon\xi_1({\bf r})+
\epsilon^2\xi_2({\bf r})+O(\epsilon^3)$ is the $\epsilon$-expansion
of $\xi$ in terms of the $\overline{n}_{\alpha}$ and $z$
(for the sake of brevity we omit the expressions of
$\xi_1$ and $\xi_2$).  By overline we are denoting weighted densities
of only the large component.

We can now insert the expansion of $\xi$ into the definition of
$\Upsilon$ [Eq.\ (\ref{upsilon})] and expand the whole functional
in powers of $\epsilon$. This leads to
\begin{eqnarray}
\Upsilon(\mu_S,[\rho_L])&=&-\Pi_0(\epsilon)V+\mu_0(\epsilon)N+
F_{\rm PAHC}+O(\epsilon) , 
\label{Fasymp}  \\
\beta F_{\rm PAHC}&=&\int d{\bf r}\,\{\overline{\Phi}_{\rm id}+
  \overline{\Phi}_{\rm ex}+\Phi_{\rm ad}\} .
\label{FPAHC}
\end{eqnarray}
The term $-\Pi_0V+\mu_0N$ diverges with $\epsilon\to 0$; nevertheless
it is irrelevant for the phase behaviour because 
it simply adds $\Pi_0$ to the pressure and $\mu_0$ to the
chemical potential of the large particles. These are two divergent
constants which just cancel out in the equilibrium equations.
The functions $\overline{\Phi}_{\rm id}$ and $\overline{\Phi}_{\rm ex}$
are given by (\ref{ideal}) and (\ref{excess}) but only for the large
component, and $\Phi_{\rm ad}$ is given by 
\begin{equation}
\Phi_{\rm ad}=\frac{z}{8}\frac{|\nabla\overline{n}_3|^2-
4\overline{\bf n}_2\cdot\overline{\bf n}_2}{1-\overline{n}_3} .
\label{Phiad}
\end{equation}

A few things are worth noticing here. First of all we can
see that out of this procedure a new weighted density has come
up---namely $\nabla\overline{n}_3$---which
was missing at the beginning. Secondly,
the second functional derivative of the interaction part of
$F_{\rm PAHC}$ yields the DCF (\ref{PAHC-DCF}), and this provides
a consistency test. Finally, $\Phi_{\rm ad}$ is a negative
contribution---it arises from an attractive interaction,
as it can be easily realized particularizing for the uniform fluid (it
can also be checked that this term is negative for {\em any} density
profile).

The equation of state of the PAHC uniform fluid, in terms of 
$y=\eta/(1-\eta)$, turns out to be
\begin{equation}
\beta P=y+3(1-z/2)y^2+2y^3.
\label{eos}
\end{equation}
As in the case of the adhesive hard spheres (AHS) fluid \cite{baxter}
this equation has a van der Waals loop, and hence yields a gas-liquid
phase transition (see Fig.\ 1) with a critical point at
$z_c=2(1+\sqrt{2/3})\approx 3.63$,
$\eta_c=(1+\sqrt{6})^{-1}\approx 0.29$.

\begin{figure}
\epsfig{file=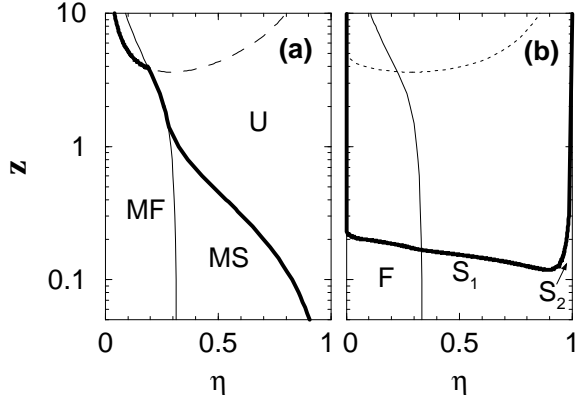, width=3.0in}
\caption{Solvent fugacity, $z$, vs.\ solute packing fraction, $\eta$,
of the infinitely asymmetric binary mixture of PHC, both without (a)
and with (b) polydispersity ($\Delta\sigma/\sigma=4.5\%$).
(a) The thick solid line separates the unstable region (U) from the 
metastable one; the thin one marks the (continuous) transition
from a metastable fluid (MF) to a metastable solid (MS); the dashed
one is the fluid-fluid spinodal. 
(b) The thick solid line marks the fluid-solid or solid-solid
coexistence; the thin one marks again the (continuous) fluid-solid
transition below the coexistence region; the dotted line is the
metastable fluid-fluid coexistence.}
\end{figure}

Allowing for
inhomogeneous phases, however, changes dramatically the phase behaviour.
As usual in density functional theories we determine the solid free
energy by writing $\rho_L({\bf r})$ as
a sum of gaussians centered at the nodes of a
(simple cubic) lattice, leaving their width as a variational parameter.
We also account for vacancies by multiplying by an occupancy probability
$\nu$, which acts as a second variational parameter. This occupancy
can be expressed in terms of the lattice parameter, $d$, as
$\nu=\eta d^3$; hence $d$ can be treated as an alternative variational
parameter---subject to the constraint $d\geq 1$. First thing we can see 
is that the fluid-solid transition is continuous---therefore we actually
obtain it directly from the DCF (\ref{PAHC-DCF}) by a standard
bifurcation analysis---and that it crosses the
fluid-fluid coexistence at a packing fraction $\eta<\eta_c$. 
Hence the fluid-fluid transition is preempted by freezing.

But having $d$ as a variational parameter reveals that any of these
phases is metastable. The reason is that {\em at any value of} $z$,
$\Phi_{\rm ad}\to -\infty$ as $d\to 1$. Thus
the only thermodynamically stable phases this system
possesses are a closed-packed solid coexisting with a zero-density fluid.
This is not surprising if one compares with the phase behaviour of
adhesive hard spheres (AHS). Computer simulations of a square-well 
fluid with well-range going to zero \cite{bolhuis}
reveal that the AHS fluid also
has the same phase behaviour. This singularity was already noticed 
by Stell \cite{stell}, who proved that the partition function of
the AHS fluid diverges for a system of 12 or more particles (12 is
the coordination number of a fcc lattice). It simply expresses the
fact that at any density and adhesiveness (solvent fugacity in
our case) the system collapses into a close packed solid lattice.

Nevertheless, a more detailed exploration of the phase diagram shows
that the fluid or the normal solid are sometimes local minima of the
free-energy functional, and thus they are metastable phases, mostly
separated from the close-packed solid by large free-energy barriers.
The limit of mechanical stability of these metastable phases is shown
in Fig.\ 1(a). It can be seen that the larger $z$ the larger the
density range of metastability.

Thus the situation one will typically find in the PAHC fluid is the
following. At large values of $z$ the fluid will quickly
collapse into the close-packed configuration.
At small values of $z$, however,
the fluid will be trapped in metastable phases for very long times; 
upon increasing $\eta$ the fluid will be seen to undergo a continuous
fluid-solid transition at $\eta\approx 0.3$, and it can remain
solid for a very long period of time; if we move at still higher
$\eta$ the system will eventually collapse, but as it may take long
for the whole system to do it there may be an apparent expanded
solid-collapsed solid coexistence. This picture is a
caricature of the typical phase diagram of a fluid exhibiting an
isostructural solid-solid transition \cite{bolhuis,tejeroetal}.

A similar behaviour has been observed
for the AHS fluid \cite{tejero}: the phase diagram reveals a freezing
transition in spite that the free energy becomes concave beyond a
certain packing fraction (depending on the adhesiveness). The latter
was interpreted in \onlinecite{tejero} as a percolation transition; 
however it is simply a sign of the collapse singularity, which did not
show up because the density of vacancies was fixed to 0 (and hence the
lattice parameter is forced to be $d=\eta^{-1/3}>1$; we observe the
same behaviour in our system when we fix $\nu=1$). The reported
phase diagram is thus metastable (compare the qualitative agreement
with the one shown in Fig.\ 1(a) for the PAHC fluid).

The collapse of the adhesive potential can be removed by adding a
small amount of polydispersity \cite{stell,frenkel}. To see what the
phase diagram looks like we have introduced
a small amount of size polydispersity
in the large cubes. In the fundamental measure
formalism this simply amounts to replacing the weights
$\omega^{(\alpha)}_L$ by their averages $\widetilde{\omega}^{(\alpha)}_L$
over the chosen size distribution. We have made the simplest computational
choice for the latter: we have replaced the large cubes by
parallelepipeds where each of their edge-lengths is chosen random and
independently from a Gaussian distribution of mean 1 and deviation
$\Delta\sigma$. It turns out that this choice transforms the former
weights into ``smoothed'' versions of them, and that this removes the
divergence of $\Phi_{\rm ad}$ no matter how small $\Delta\sigma$ be.
Besides, this choice leaves the free energy of the uniform fluid
unchanged.

The results are plotted in Fig.\ 1(b).
It can be seen that now we really have the typical phase
diagram of a fluid exhibiting an isostructural solid-solid
transition---except that for this system the fluid-solid transition
is continuous. For large values of $z$ we have a very diluted fluid
and a very dense solid separated by a large coexistence region. As we
decrease $z$ the coexisting fluid eventually becomes denser until
it reaches the density at which freezing occurs. From then on and down
to a critical value, $z_s$, coexistence is between
an expanded and a dense solid with the same crystal structure.

The results we have obtained can be reinterpreted in terms of the
original very asymmetric binary mixture. The fluid phase would
correspond to the stable mixture, and the solid phases to the
solid of large particles permeated by a fluid of small particles.
According to the PAHC phase diagram---either mono or polydisperse---no
fluid-fluid phase separation will ever occur for very asymmetric 
binary mixtures of PHC; instead, at large solvent fugacity (large
solvent density) there will be fluid-solid phase separation, and at
smaller solvent fugacity the phase separation will take place {\em
after} the large particles have crystallized. 

Notice that the existence of this fluid-solid phase separation
largely increases the region of the phase diagram where the mixture
is unstable. This explains why recent simulations on very
asymmetric binary mixtures of PHC \cite{buhot} report the mixture
to be unstable at packing fractions of the large component much 
smaller than those predicted by fluid-fluid phase separation
\cite{cuesta} (notice that, according to the authors, the nature of
the coexisting phases cannot be discerned in their simulation).

The same kind of phase behaviour has been very recently reported to
occur in simulations of binary mixtures of HS at finite (but small)
diameter ratios using the depletion potential \cite{noe,dijkstra}
(they also show a
solid-solid transition in a certain range of solvent fugacities).
We can see then that this phase behaviour can be inferred from that
of the AHS fluid \cite{tejero}. This result is of great importance
because it establishes a connection between the phase behaviour of 
mixtures and that of the solvent fluid interacting via an adhesive-like
potential \cite{bolhuis,tejeroetal}, thus opening a new route to
study demixing in hard-particle fluids. This remarkable enhancement
of the slow divergence of the pair correlation function of two large
bodies at contact, induced by depletion \cite{biben}, is probably
due to the fact that the confinement of the solvent in between the
solute particles induces strong inhomogeneities in the former 
[see Eq.\ (\ref{equil})] an hence largely increase depletion.

It is a pleasure to thank Daan Frenkel and Pedro Tarazona
for illuminating discussions.
This work is supported by project no.\ PB96-0119 from the
Direcci\'on General de Ense\~nanza Superior (Spain).


\end{multicols}

\end{document}